# Growth Direction Dependence of the Electron Spin Dynamics in {111} GaAs Quantum Wells

H.Q. Ye[1], G. Wang[1], B.L. Liu[1,a)], Z.W. Shi[1], W.X. Wang[1], C. Fontaine[2], A. Balocchi[3], T. Amand[3], D. Lagarde[3], P. Renucci[3], X. Marie[3,b)]

[1]Beijing National Laboratory for Condensed Matter Physics, Institute of Physics, Chinese Academy of Sciences, P.O. Box 603, Beijing 100190, China

[2]CNRS, Université de Toulouse, LAAS, 7 avenue du Colonel Roche, F-31400 Toulouse, France

[3]Université de Toulouse, INSA-CNRS-UPS, LPCNO, 135 avenue de Rangueil, 31077 Toulouse, France

## Abstract

The electron spin dynamics is studied by time-resolved Kerr rotation in GaAs/AlGaAs quantum wells embedded in a negatively doped-intrinsic-positively doped structures grown on (111)A or (111)B-oriented substrates. In both cases the spin lifetimes are significantly increased by applying an external electric field but this field has to point along the growth direction for structures grown on (111)A and opposite to it for the ones grown on (111)B. This extended electron spin lifetime is the result of the suppression of the D'yakonov-Perel spin relaxation mechanism [Sov. Phys. Solid State **13**, 3023 (1972)] due to the cancellation effect of the internal Dresselhaus term [Phys. Rev. **100**, 580 (1955)] with the external electric field induced Rashba one [J. Phys. C **17,** 6039 (1984)], both governing the conduction band spin-orbit splitting. These results demonstrate the key role played by the growth direction in the design of spintronic devices.

a),b) Authors to whom correspondence should be addressed :

marie@insa-toulouse.fr

blliu@iphy.ac.cn

The electrical control of the electron spin dynamics in semiconductors is of fundamental importance for many potential applications in spintronics [1-3]. Due to the combined effect of spin orbit interaction and the absence of symmetry inversion, the dominant electron spin relaxation mechanism in III-V or II-VI zinc-blende semiconductor quantum wells (QW), called the D'yakonov-Perel (DP) mechanism, depends both on the conduction band (CB) spin-orbit splitting and the electron wavevector relaxation time. The DP spin relaxation time $\tau_s$ writes simply $\tau_s^{-1} = \langle \Omega^2 \rangle \tau^*$, where $\langle \Omega^2 \rangle$ is the mean square of $\mathbf{\Omega}$, the effective magnetic field corresponding to the CB spin-splitting and $\tau^*$ the electron momentum relaxation time [4]. The CB spin splitting is mainly due to two additive contributions: the Bulk Inversion Asymmetry (BIA or Dresselhaus contribution [5]), and the Structural Inversion Asymmetry (SIA or Rashba contribution [6]). The latter can be induced by applying an external electric field to the nanostructure. As a consequence, the electron spin relaxation time in quantum wells can in principle be tuned by the control of an electric field induced by an external voltage or an asymmetric doping in the structure [7].

There have been many theoretical and experimental works in the past few years on the control of the electron spin relaxation based on the choice of specific crystal orientations and appropriate relative strengths of the Rashba and Dresselhaus terms [8-16]. It was demonstrated that the DP electron spin relaxation can be significantly increased, and in some cases suppressed for one (and only one) given electron spin direction either in (001) or (110) quantum wells. In (001) GaAs quantum wells longer DP spin relaxation times along the in-plane [110] direction were demonstrated as a result of the interplay between BIA and SIA terms [17–19] ; for electrons spins aligned along other directions, very efficient spin relaxation occurs. For (110) quantum wells, it turns out that the effective magnetic field due to the BIA term is oriented along the [110] growth direction [4]; if the electron spin is also aligned along this direction, the DP spin relaxation mechanism is suppressed but electron spins aligned along other directions experience fast relaxation [20,21].

In {111} quantum wells, setting the **z**-axis along the [111] direction, the **x**-axis along [11–2] and the **y**-axis along [–110], the Dresselhaus first order Hamiltonian for the first electron sub-band writes [4] :

$$H_D = \beta(k_y \sigma_x - k_x \sigma_y), \qquad (1)$$

where $\sigma_i$ ($i = x,y,z$) are the Pauli spin matrices, and $\beta = \frac{2\gamma\langle\hat{k}_z^2\rangle}{\sqrt{3}}$, with $\gamma$ the Dresselhaus coefficient of the QW structure and $\langle\hat{k}_z^2\rangle$ the averaged square of the electron linear momentum component along the growth direction (proportional to the electron confinement energy). In {111} QWs it turns out that the Rashba Hamiltonian for an applied electric field along the growth direction **z** has exactly the same form as the Dresselhaus one [22-24]; it simply writes:

$$H_R = \alpha(k_y\sigma_x - k_x\sigma_y) ,\qquad(2)$$

where $\alpha = a.E$ is the Rashba coefficient and $E$ the electric field. The combination of Eq. (1) and (2) yields the total spin-orbit coupling Hamiltonian for {111} QW:

$$H = (\alpha + \beta)(k_y\sigma_x - k_x\sigma_y) \qquad(3)$$

When $\alpha = -\beta$, the CB spin-orbit splitting vanishes and one expects an overall suppression of the DP spin relaxation mechanism for all the 3 spatial orientations of the CB electron spin [22]. This theoretical prediction was recently verified experimentally by applying an external electric field in GaAs/AlGaAs quantum wells embedded in a *NIP* structure grown on (111)B substrates [25]. The electron spin relaxation time as measured by time-resolved photoluminescence can be increased by more than two orders of magnitude reaching values larger than 30ns.

In this Letter we demonstrate that the direction of the external electric field required to suppress the electron spin relaxation in {111} quantum well structures depends on the substrate orientation (111)A or (111)B (Gallium rich and Arsenic rich respectively). We find that the required electric field has to point along the growth direction for structures grown on (111)A and opposite to it for those grown on (111)B. We show here that in addition to the well-known tuning of the Dresselhaus term through the confinement and the tuning of Rashba term through the electric field amplitude, the growth direction represents another degree of freedom which can prove an efficient engineering tool for spin relaxation times in quantum wells.

In this work we have studied four samples grown by molecular beam epitaxy. One *NIP* (*PIN*) structure grown on (111)A GaAs substrate and one *NIP* (*PIN*) structure grown on (111)B substrate (figure 1). The (111)B substrates are misoriented 3° towards <111>A and the orientation of (111)A substrates are (111)A ± 0.5°. The layers sequence of *NIP* structure is: substrate/200nm GaAs buffer/ 500nm p-doped ($p = 1.2 \times 10^{18}$ cm$^{-3}$) GaAs/ 100nm Al$_{0.3}$Ga$_{0.7}$As barrier/MQW/70nm Al$_{0.3}$Ga$_{0.7}$As barrier/ 10nm n-doped ($n = 2.8 \times 10^{18}$ cm$^{-3}$) GaAs (see figure 1 for the two *NIP* structures). In the PIN samples the *n*- and *p*-doped layers are reversed in the sequence. The Multiple Quantum Well (MQW) layer consists of 20 periods GaAs/Al$_{0.3}$Ga$_{0.7}$As quantum wells with $L_W$ = 12 nm well width and $L_B$ =30nm barrier width for (111)A samples and $L_W$ = 15 nm and $L_B$ =12 nm for (111)B samples.

The electron spin dynamics has been studied by Time Resolved Kerr Rotation experiments [17]. The sample are excited near normal incidence with degenerate pump and delayed probe pulses by a Ti:Sapphire laser (pulse width :120 fs and repetition frequency : 76 MHz). The pump beam is circularly polarized and the probe beam is linearly polarized with an average power of 10 and 0.5 mW respectively. Complementary measurements of the electron spin dynamics were obtained from the time and polarization resolved photoluminescence kinetics using circularly polarized picosecond Ti: Sapphire laser pulses and a synchroscan Streak camera with an overall time-resolution of 8 ps [27]. All the experiments presented here have been performed at *T*=50 K ; similar results were obtained up to *T*=150 K on the different samples and confirm the data at *T*=50 K.

Figure 2(a) shows the time evolution of the Kerr rotation signal for different applied voltages in the *NIP* structure grown on (111)B at *T*=50K. The electron spin lifetimes $T_s$ can be extracted by a linear fit in logarithmic coordinates. We observe a clear increase of $T_s$ when the reverse bias increases: $T_s$ *increases* from about 380 ps to 830 ps when the gate voltage varies from +0.4 V to –2.6 V (inset of figure 2(a)). The spin lifetime measured in the TRKR depends both on the electron spin relaxation time and the carrier lifetime $\tau_r$ : $1/T_s = 1/\tau_s + 1/\tau_r$. From the measurement of $\tau_r$ by time-resolved photoluminescence, we can extract the electron spin relaxation time (for instance at *V*=–2.6 V, we find $\tau_r$ ~ 3200 ps yielding an electron spin relaxation time $\tau_s$ ~ 1100 ps, in agreement with previous measurements for this applied electric field [25]). From this analysis we conclude that the large increase of $T_s$ is the consequence of the strong augmentation of the electron spin relaxation time $\tau_s$. The latter effect is due to the reduction of the total CB electron spin-splitting induced by the Rashba term when it has opposite sign but same direction compared to the Dresselhaus one (see equation 3) [28].

Figure 2(b) displays the result of the same experiment performed on the *NIP* structure grown on the (111)A substrate. The striking feature is that applying a reverse bias in this sample yields *a decrease* of the electron spin lifetime, i.e. exactly the opposite behavior compared to the (111)B sample. When the voltage varies from + 3V to –1V, the spin lifetime *decreases* from ~ 310 ps to 140 ps (inset of figure 2(b)). These results are confirmed by time and polarization resolved photoluminescence experiments displayed in figure 3. The decay time of the circular polarization of the photoluminescence corresponds directly to the electron spin relaxation time $\tau_s$ which drops drastically when the voltage varies from +2V to –5V, in agreement with the TRKR measurements .

As shown in the I-V curve measurements in figure 1(c), $V \sim +3$ volts corresponds approximately to the flat band conditions, i.e. the electric field in the MQW is almost zero and the electron spin relaxation time is only governed by the Dresselhaus term. For the other bias voltages, the electric field in the MQW region can be determined from the measurement of the Quantum Confined Stark Effect (QCSE) [29] and the values are in agreement with the dimension of the structure ; the inset in figure 3 displays the red-shift corresponding to QCSE of the photoluminescence peak energy when the reverse bias increases. For $V = -3$ volts, a red shift of about 14 meV is observed (with respect to the flat band situation) corresponding to an electric field amplitude of $|E| \sim 70$ KV/cm.

We note in figure 2 that the spin lifetimes measured for $V \sim +2$ volts (almost zero external field in the structures) is shorter in the (111)A sample ($T_s \sim 260$ ps) compared to the value ($T_s \sim 400$ ps) measured in the (111)B sample. The spin relaxation times in the two structures cannot be directly compared since the quantum well widths are different in the two samples ($L_W = 15$ nm for (111)B and $L_W = 12$ nm for (111)A samples). As the Dresselhaus term is proportional to $\langle \hat{k}_z^2 \rangle \approx (\pi / L_W)^2$, it is about 50% larger in the $L_W = 12$ nm structure yielding a shorter spin relaxation time. As a consequence it is more difficult to compensate the Dresselhaus spin splitting by the electric field induced Rashba term in the (111)A sample compared to the (111)B one.

A simple illustration, shown in figure 4, summarizes the experimental results. We set the **z**-axis along the direction [111] and fix it, whatever the sample is. For the (111)B *NIP* sample, the observation of an extended spin lifetime requires the application of a reverse bias, *i.e.* the electrons in the quantum wells feel an electric field **E** opposite to the growth direction, that is **E** is oriented along the **z** axis. Using the same coordinate axis, our experimental results on the (111)A *NIP* sample show that the electron spin relaxation time *decreases* if the electric field amplitude, still oriented along z, increases (reverse bias). In other words, this means that the increase of spin relaxation time in the QWs grown on (111)A substrates also requires the application of an external electric field along the **z** direction, which is in this case the growth direction.

In order to check the reversal of the external electric field orientation (with respect to the growth direction) required to increase the electron spin relaxation time when a structure is grown on (111)B or (111)A substrates, we have also investigated two *PIN* structures grown on these two kinds of substrates. In reverse bias, the electric field orientation in the *PIN* structure is opposite to the one in the *NIP*. As expected we find the opposite behavior compared to the one measured in figure 2 for *NIP* samples, i.e. the electron spin relaxation time *decreases* with the reverse bias in the *PIN* (111)B structure and *increases* in the (111)A *PIN* sample (not shown).

We emphasize that although the applied electric field **E** has to be reversed with respect to the growth direction for samples grown on the different substrate orientation (111)A or (111)B, the vector **E** required to achieve the spin splitting cancellation must always be oriented along the fixed crystallographic axis [111] as shown in figure 4. It is simply the consequence of the relative sign of the Rashba and Dresselhaus coefficients, which can be calculated within the 14 bands **k.p** model : this sign is a characteristic of the material pair used to grow the heterostructure [7,30]. From the point of view of the crystallographic axis, the results displayed in figure 2a and 2b correspond to the same physical situation since the positive and negative voltages are defined with respect to the standard diode convention and not to the fixed crystallographic orientation.

In conclusion, we have studied the electron spin dynamics in {111}-oriented GaAs *NIP* and *PIN* quantum wells devices grown on (111)A or (111)B substrates. By applying an external electric field, the spin lifetimes can be significantly increased due to the cancellation effect of Rashba and Dresselhaus terms. The increased spin lifetime is observed if the applied external electric field points *along* the growth direction for structures grown on (111)A and *opposite* to it for the ones grown on (111)B. This result is important for the design of devices based on the unique spin properties of (111)-oriented nanostructures.

Acknowledgments:

This work was supported by the France-China NSFC-ANR research project SPINMAN (Grant No. 10911130356), CAS Grant No. 2011T1J37, National Basic Research Program of China (2009CB930500) and National Science Foundation of China (grant number 10774183, 10874212); we thank H. Carrère and A. Arnoult for useful discussions and X. D. Wang and A. Z. Jin for technical support.

**Figure Captions :**

Figure 1

Schematics of *NIP* GaAs/AlGaAs quantum wells structure grown on (a) (111)B and (b) (111)A substrate. The voltage is positive when the diode is forward biased.

(c) *I-V* characteristics of the *NIP* structure grown on (111)A substrate at T=50 K

(d) Simplified band structure of the NIP device with the electric field E~($V_B$+V)/d, where $V_B$ is the built-in electric field and V the applied bias.

Figure 2

Kerr rotation dynamics at *T* = 50K for *NIP* samples grown on (a) (111)B and (b) (111)A substrates with different applied voltages. The symbols correspond to the experimental results and the full line to mono-exponential fits. Insets in (a) and (b): dependence of the electron spin lifetime with the applied voltage.

Figure 3

Photoluminescence (PL) circular polarization dynamics for three applied voltages at T=50 K in the *NIP* structure grown on (111) A. The excitation laser energy is 1.563 eV and the detection energy corresponds to the PL peak displayed in the inset as a function of the applied voltage. The inset displays the PL peak shift as a function of the applied bias, resulting from quantum confined Stark effect.

Figure 4

Schematic representation of the crystal structure of the *NIP* samples grown on (111)A and (111)B substrates. The applied electric field **E** is represented here with the direction required to observe an *increase* of the electron spin relaxation time.

**Figure 1**

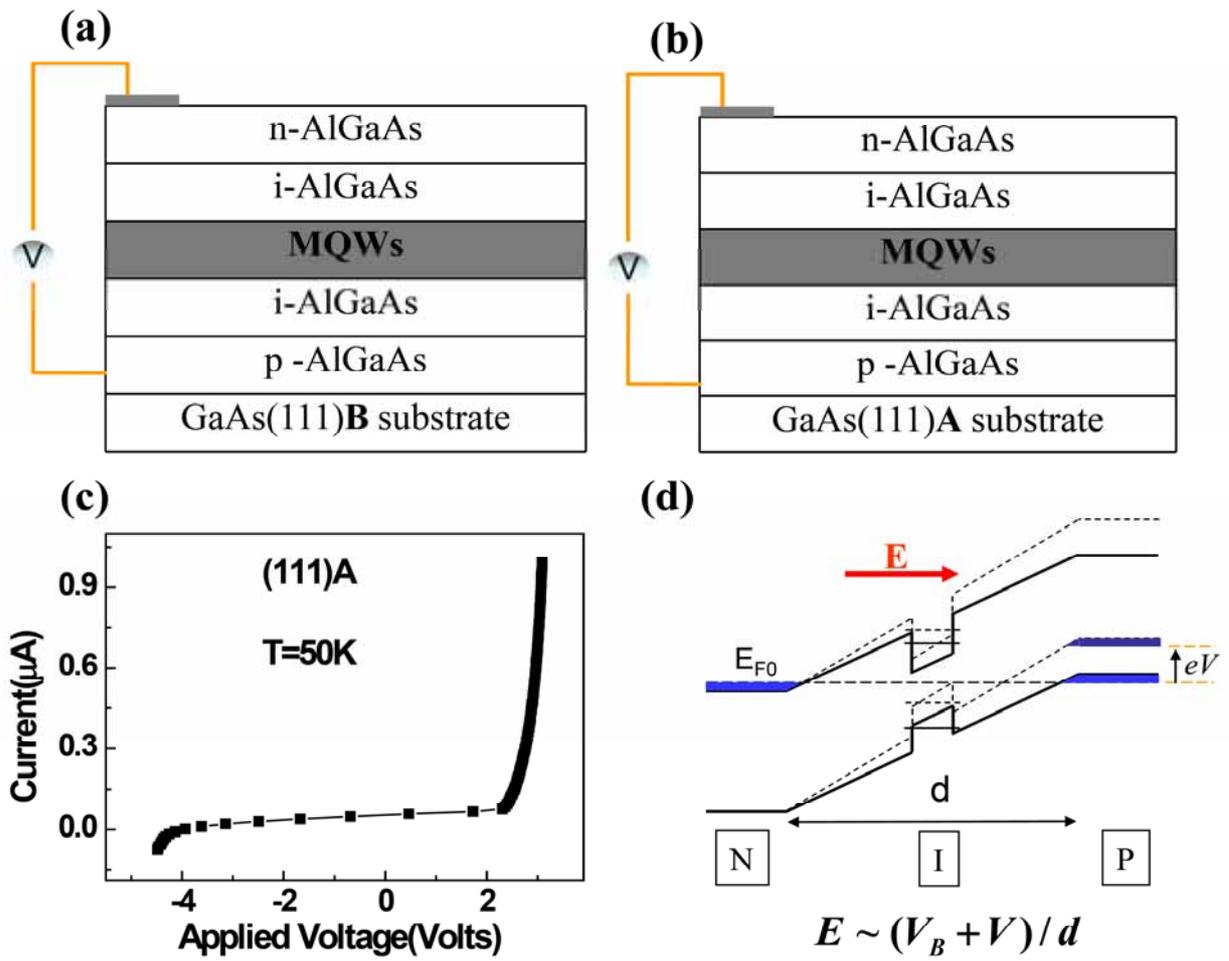

**Figure 2**

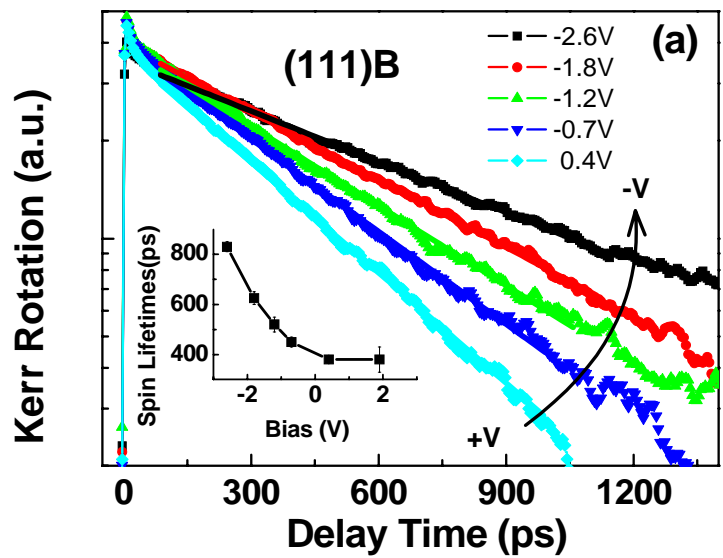

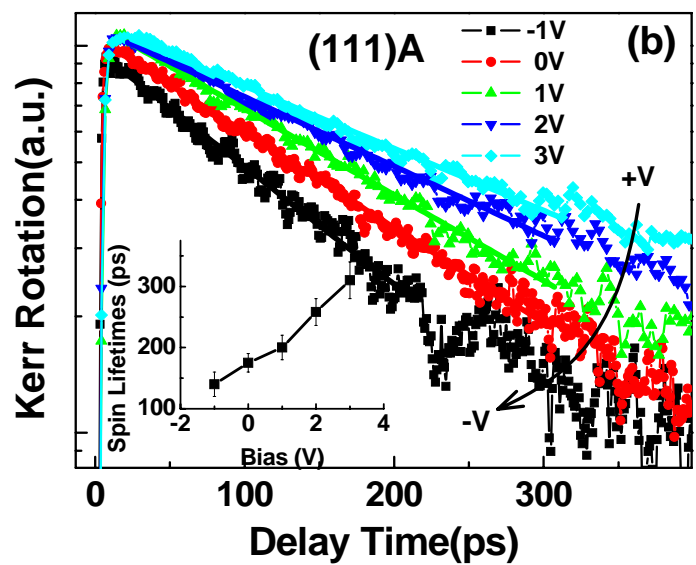

**Figure 3**

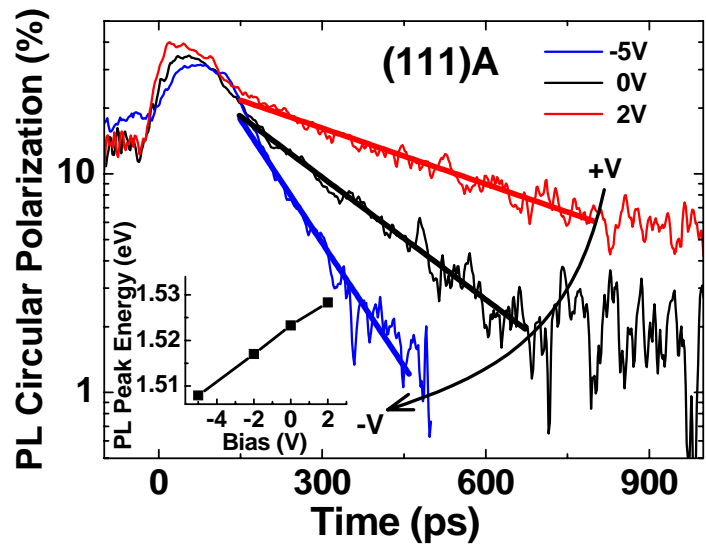

**Figure 4**

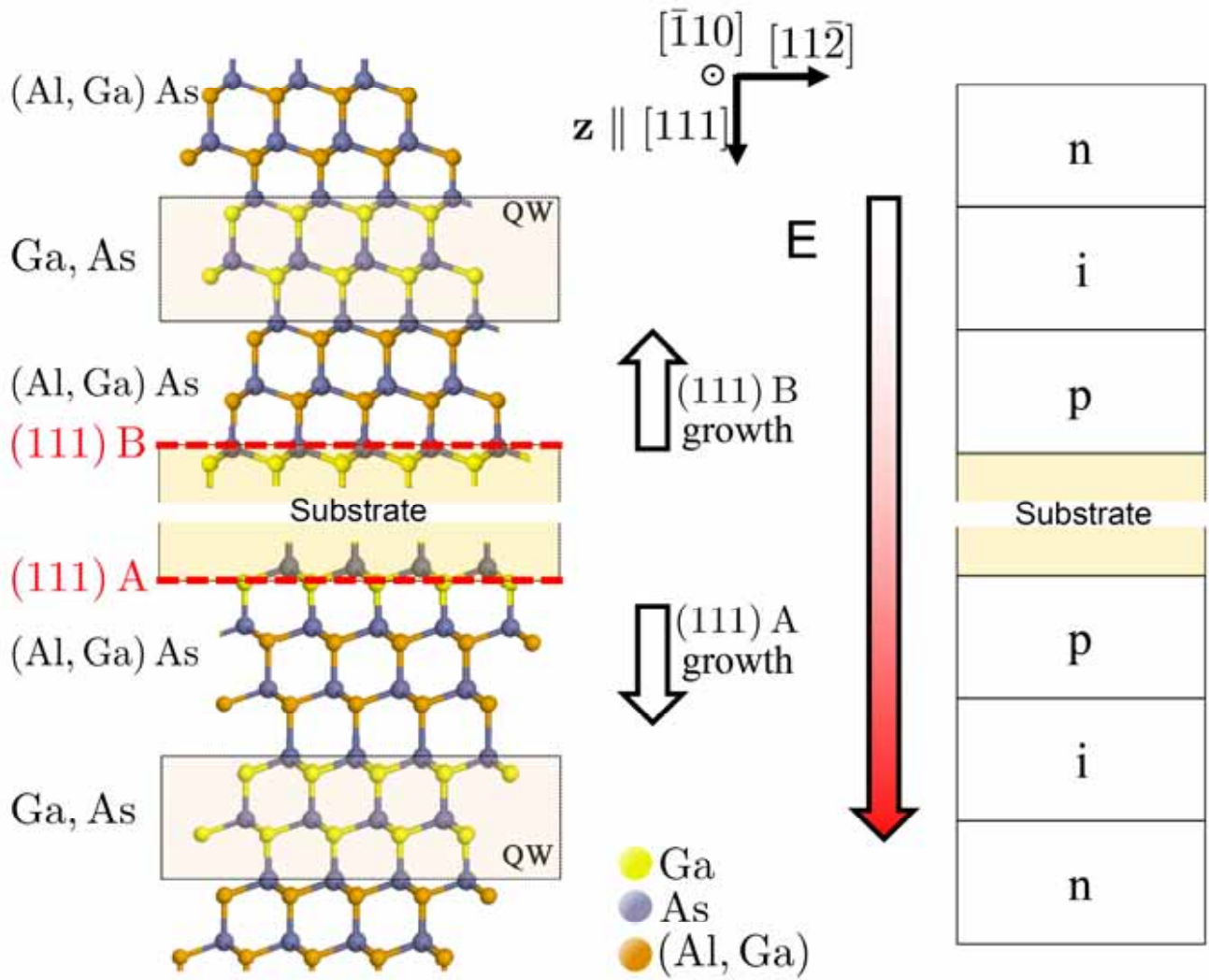